\documentclass[sigconf]{acmart}

\AtBeginDocument{%
  \providecommand\BibTeX{{%
    \normalfont B\kern-0.5em{\scshape i\kern-0.25em b}\kern-0.8em\TeX}}}

\copyrightyear{2025}
\acmYear{2025}
\acmConference[UIST '25]{The 38th Annual ACM Symposium on User Interface Software and Technology}{September 28-October 1, 2025}{Busan, Republic of Korea}
\acmBooktitle{The 38th Annual ACM Symposium on User Interface Software and Technology (UIST '25), September 28-October 1, 2025, Busan, Republic of Korea}\acmDOI{10.1145/3746059.3747770}
\acmISBN{979-8-4007-2037-6/2025/09}

\def\sys{\textsc{ProMemAssist\xspace}}

\newcommand{\revise}[2][black]{\textcolor{#1}{#2}}

\usepackage{makecell}
\usepackage{longtable}
\usepackage{listings}
\usepackage{xcolor}
\usepackage{array}
\usepackage{amsmath}
\usepackage{microtype}
\usepackage{graphicx}
\usepackage{float}
\usepackage{wrapfig}
\usepackage{xspace}
\usepackage{enumitem}
\setlistdepth{9}

\usepackage{lipsum}
\usepackage{todonotes}
\usepackage{tikz}
\begin{document}

% \title[\mk{\projname{}}]{\mk{Long Paper Title}}
\title[\sys: Working Memory Modeling for Timely Proactive Assistance]{\sys: Exploring Timely Proactive Assistance Through Working Memory Modeling in Multi-Modal Wearable Devices}

% \author{ABC DEF} 
%     \affiliation{%
%     \institution{Institute for Computer Science \\ University of Duisburg-Essen}
%             \streetaddress{}
%             \city{Essen}
%             % \state{}
%             \country{Germany}
%             \postcode{45141}
%     }

\author{Kevin Pu}
\authornote{Project completed during an internship at Meta Reality Labs.}
\email{jpu@dgp.toronto.edu}
\affiliation{
    \institution{University of Toronto}
    \country{Canada}
}
\author{Ting Zhang}
\email{tingzhang@meta.com}
\affiliation{
    \institution{Meta Reality Labs}
    \country{USA}
}
\author{Naveen Sendhilnathan}
\email{naveensn@meta.com}
\affiliation{
    \institution{Meta Reality Labs}
    \country{USA}
}
\author{Sebastian Freitag}
\email{freitag@meta.com}
\affiliation{
    \institution{Meta Reality Labs}
    \country{USA}
}
\author{Raj Sodhi}
\email{rsodhi@meta.com}
\affiliation{
    \institution{Meta Reality Labs}
    \country{USA}
}
\author{Tanya Jonker}
\email{tanya.jonker@meta.com}
\affiliation{
    \institution{Meta Reality Labs}
    \country{USA}
}

\renewcommand{\shortauthors}{Pu et al.}
\begin{abstract} 
    Wearable AI systems aim to provide timely assistance in daily life, but existing approaches often rely on user initiation or predefined task knowledge, neglecting users' current mental states. We introduce \sys{}, a smart glasses system that models a user's working memory (WM) in real-time using multi-modal sensor signals. Grounded in cognitive theories of WM, our system represents perceived information as memory items and episodes with encoding mechanisms, such as displacement and interference. This WM model informs a timing predictor that balances the value of assistance with the cost of interruption. In a user study with 12 participants completing cognitively demanding tasks, \sys{} delivered more selective assistance and received higher engagement compared to an LLM baseline system. Qualitative feedback highlights the benefits of WM modeling for nuanced, context-sensitive support, offering design implications for more attentive and user-aware proactive agents.
\end{abstract}


% \begin{teaserfigure}
%   \centering
%   % \vspace{-3mm}
%   \includegraphics[width=\textwidth]{figures/Teaser_Figure_2.png}
%   % \missingfigure{Teaser figure}
%   % \vspace{-7mm}
%   \caption{
%   While many prior systems focused on helping users explore a set of diverse ideas broadly, we present \projname{}, an ideation tool that helps users more deeply develop an initial research idea into a concrete and specific research brief. Intuitively, researchers often dissect research projects into different facets to further develop them; \projname{} also allows users to structure their idea as faceted idea nodes, explore and evolve different variations, and, finally, compose them into a complete research brief.
%   }
%   \label{fig:teaser}
% \end{teaserfigure}

\maketitle

\section{Introduction}

Context-aware wearable AI devices such as smart glasses \cite{RaybanMetaGlasses}, necklaces \cite{FriendAINecklace}, and pins \cite{humaneAIPin} are beginning to reshape how intelligent assistants support users in daily life. Their hands-free, always-on form factor enables access to real-time sensor data and provides the opportunity to proactively assist users in dynamic, situated contexts — from cooking and organizing to planning or navigating physical spaces \cite{arakawa_prism-observer_2024, baraglia_initiative_2016, cha_hello_2020}. As these systems move beyond desktop and mobile environments, a central challenge emerges: when should they step in to help?

Many existing assistants rely on user-initiated interaction, such as voice commands or manual input. % TODO: Cite
While effective in many contexts, this model assumes that users are aware of what help is possible and cognitively available to ask for it. However, in everyday tasks, users are often mentally occupied or physically engaged, making it difficult to initiate help-seeking even when assistance would be beneficial. More critically, users may not recognize when support is relevant — or may simply forget to ask.

Recent works in proactive assistants attempt to address this issue by triggering assistance based on task context or rule-based heuristics \cite{yang_amma_2024, arakawa_prism-observer_2024, li_satori_2024, liu_compeer_2024, hu_designing_2024}. For example, systems like PrISM-Observer \cite{arakawa_prism-observer_2024} and Satori \cite{li_satori_2024} leverage task step detection or inferred user goals to display timely instructions or reminders. However, these strategies often rely on predefined task structures or heuristic-based triggers. They are limited in their ability to account for internal mental states, such as attention, focus, or cognitive load, which play a critical role in determining whether a user is ready or receptive to assistance. Without such awareness, proactive support risks becoming mistimed, disruptive, or even ignored.

In this paper, we propose a novel approach to inform the timing of assistance: modeling the user’s working memory (WM) as a lens into their mental availability and the assistance's value. WM is a cognitive system responsible for temporarily holding and manipulating information during task performance \cite{baddeley_working_2012, cowan_magical_2010,byrne_computational_1996}. By modeling the contents and constraints of WM — including capacity limits, recency of information, and susceptibility to interference — we can better infer moments when users are more cognitively open to external input, and conversely, when interruptions may be costly.

We introduce \sys{}, a smart glasses system that leverages multi-modal sensor signals (camera and microphone) to construct a real-time model of a user’s working memory. The system encodes visuospatial and phonological memory items from the environment and binds them into episodic chunks that represent meaningful task contexts \cite{baddeley_working_2012,cowan_magical_2010,miller_magical_1956}. A timing predictor uses this WM model to evaluate the potential value of assistance and the cognitive cost of interruption, selecting moments when support is most likely to be helpful and least disruptive. Assistance is generated using a large language model (LLM) based on current memory state. 
% The focus our work is on the timing of delivering proactive assistance, thus, we are not emphasizing on assistance generation or goal inference. 
The focus of our work is on \textit{when} assistance should be delivered, rather than \textit{what} assistance to provide. As such, our system does not aim to optimize goal inference and the content of assistance beyond what can be reasonably inferred from the current working memory state.

To evaluate our approach, we conducted a within-subject user study with 12 participants performing real-world tasks that demand both physical interaction and cognitive engagement. Participants completed four tasks (e.g., setting up a dining table, packing for a trip) while receiving assistance either from \sys{} or a baseline system where an LLM agent with similar system prompts dictated the timing and generation of support based on the same environmental information. Our findings show that \sys{} delivered assistance more selectively depending on \revise{the user's WM model}, with richer positive engagement and fewer negative reactions.

This paper makes the following contributions:
\begin{itemize}
% [leftmargin=*]
    \item A novel working memory modeling framework for determining opportune moments to deliver proactive assistance on wearable devices;
    % \item A system architecture that combines multimodal perception, WM encoding, and LLM-based intervention generation;
    \item A proactive timing prediction approach that balances the value and cost of assistance based on cognitive state;
    \item A user study demonstrating that WM-informed timing leads to improved user experience and engagement.
\end{itemize}
\section{Related Work}

% \subsection{Wearable Assistants}
% - Wearable system that are user-initiated
% - systems that are proactive but are using predefined task knowledge
% - systems that are using heuristic based timing
% - Briefly mention the systems that model user mental state and highlight the current gaps

% \subsection{Timing of Service and Interruptability}
% - Discuss principles of proactive timing and prior results on when to interrupt

% \subsection{Working Memory}
% - Essential cognitive psych theories that informs our system design

% \subsection{Memory-Augmentation Systems}
% - Discuss structure of working memory and existing approach to model memory

\subsection{Wearable Device for Task Assistance}
Advancements in sensing and lightweight computing devices have led to novel wearable assistants in consumer and research fields alike.
LLM-enabled commercial products like Ray-ban glasses \cite{RaybanMetaGlasses}, AI pin \cite{humaneAIPin}, and AI friend necklace \cite{FriendAINecklace} produce a plethora of possibilities and controversies regarding how technologies can assist or disrupt people's daily lives.
Recent research work has demonstrated the potential of wearable and situated agents to support users in physical environments by leveraging multi-modal input streams and AI models. 
For example, Arakawa et al. utilized smart watch sensing and task step modeling to predict the user's progress and provide just-in-time interventions \cite{arakawa_prism-observer_2024} and even answer voice queries \cite{arakawa_prism-q_2024}.
Another work, OmniActions, predicts digital follow-up actions, such as looking up information or sharing captured images, based on real-world multi-modal signals using LLMs \cite{li_omniactions_2024}. 

In VR/AR environements, works like AMMA adapt task guidance interfaces by modeling user state and planning adaptive step-by-step support \cite{yang_amma_2024}.
AdapTutAR presents in-situ task guidance by detecting user behavior through AR glasses \cite{huang2021adaptutar}.
Another work, Satori, forecasts the user's task actions by modeling their intention and present assistance in AR \cite{li_satori_2024}.
These systems illustrate progress in translating ambient perception into user task understanding and actionable system guidance.
Our work differs from these systems by focusing on \revise{constructing a model of} the user's cognitive state, specifically working memory (WM), rather than \revise{inferring perceptibility} based on external environmental or task cues. Unlike proactive support that relies on observed task progress or heuristics, we introduce a timing predictor informed by cognitive modeling, enabling interventions that are responsive to mental load and availability.

\subsection{Memory Augmentation and Modeling}

Memory augmentation has long been a research goal in ubiquitous and wearable computing. Early visionaries like Lamming and Rhodes imagined memory prostheses that could recall contextually relevant past information \cite{lamming_design_1994, rhodes_remembrance_1996}. 
% Early systems such as the Remembrance Agent \cite{rhodes_remembrance_1996} and Memoro \cite{zulfikar_memoro_2024} have explored passive memory augmentation through continuous retrieval, but lacked moment-to-moment sensitivity to cognitive cost. ProMemAssist builds on this trajectory by embedding a real-time cognitive state model to govern delivery, not just content.
More recent work such as SenseCam diaries \cite{lee_constructing_2008}, MemoriQA \cite{tran_memoriqa_2024}, and OmniQuery \cite{li_omniquery_2024} support retrospective access to personal memory for tasks like information storage, question-answering, and life-logging \cite{harvey_remembering_2016}.
In addition, many works augment user's capabilities to recall information.
For example, Memoro offers lightweight real-time augmentation with mixed-initiative memory aid \cite{zulfikar_memoro_2024}, and Shen et al. constructed a VLM-based memory augmentation agent for episodic memory recall tasks \cite{shen_encode-store-retrieve_2024}.
\revise{Another tool, AiGet, uses sensing data from AR glasses to construct knowledge library of the user's daily life to provide proactive interventions~\cite{cai2025aiget}.}
Moreover, prior work in XR memory systems explores how immersive technologies can externalize memory to improve recall and reduce load \cite{makhataeva_augmented-reality-based_2023, bonnail_memory_2023}

While these systems primarily focus on retrieval support,
\sys{} addresses a complementary but underexplored design space: timing real-time proactive support to align with ongoing mental processes. We argue that effective memory augmentation should not only improve recall but also account for the user’s cognitive readiness in the moment.
\revise{Our approach is partially grounded in the Cognitive Load Theory \cite{SWELLER201137}, which puts mental workload as a key consideration for interaction and proposes that the timing and modality of intervention could impact user mental workload and receptivity. }

\revise{One approach to address cognitive receptivity is through physiological sensing. For example, Sarker et al. developed a machine learning model to assess user availability for just-in-time interventions based on wearable sensor data (ECG, accelerometer, respiration) and reported over 74\% accuracy in natural environments \cite{Sarker2014AssessingTA}. Similarly, Chan et al. introduced Prompto, a memory training assistant that initiates prompts when electrodermal activity (EDA) and heart-rate variability (HRV) signals suggest the user is under low cognitive load \cite{Chan2020PromptoIR}.}
By contrast, \sys{} infers cognitive availability by constructing mental workload model using observable environmental cues (e.g., visual and auditory signals) rather than physiological signals. While physiological sensing provides fine-grained access to internal states, it can be intrusive, less interpretable, and harder to generalize across users. 
\sys{} offers a more lightweight, explainable model for attention dynamics based on working memory constructs (e.g., recency, interference), and complements prior work by providing an alternative path to cognitively aligned assistance.

% Our work connects these threads by introducing structured cognitive modeling into the proactive support pipeline.
% Together, these approaches point to a broader opportunity: integrating cognitive models, physiological signals, and behavioral context to create more responsive and human-aligned interventions. 
% Future work could explore hybrid systems that combine real-time WM modeling with physiological sensing to improve robustness across settings.

Another major application for memory modeling is to simulate and predict user goals or attention via memory modeling \cite{laird_it_2001, park_generative_2023, bahrainian_predicting_2019, bahrainian_towards_2017, bahranian2018augment, umarov_believable_2012}, but few systems have attempted to model the working memory explicitly for timing decisions. 
MATCHS, for example, uses WM simulation to adapt interface difficulty for older users or those with cognitive impairments \cite{massoni_sguerra_adapting_2018, sguerra2019hound}, but does not guide assistance delivery timing.

Our contribution is not to offer a definitive cognitive model that can predict user needs and behaviors, but to explore how lightweight WM modeling can inform system timing in interaction. This offers a new mechanism for designing more attentive and personalized just-in-time assistants, particularly in AR and wearable contexts where attention and cognitive load are highly volatile.

\subsection{Timing of Service and Interruptibility}

A central challenge in mixed-initiative systems is knowing when to assist \cite{horwitz1999mixedinitiative,amershi2019haiprinciple}. 
Well-timed assistance can ease cognitive effort, avoid confusion, and even foster incidental learning about the system’s capabilities \cite{matejka2011ambient, horwitz1999mixedinitiative, Sawyer_2014_learning}.
In contrast, poorly timed interventions may negatively impact users' memory, emotional well-being, and ongoing task execution~\cite{bailey2000measuring,horvitz2001notification,czerwinski2000instant}. 

Past systems have explored timing proactive assistance by modeling user behavior and task states \cite{rabbi_mybehavior_2015, liu_compeer_2024, chen_need_2024, Pu2025AssistanceOD, cha_hello_2020}.
Additionally, research on notification timing has shown that context-based deferral can improve user experience \cite{li_alert_2023, chen2022opportune}.
Efforts to model interruption cost have led to systems that predict notification acceptability \cite{meurisch_exploring_2020} or adapt service robot initiative \cite{baraglia_initiative_2016, peng_design_2019}. 
While these approaches often use task context, heuristics, or rule-based timing triggers and interruption measurement, \sys{} models WM mechanisms such as internal cognitive interference and recency decay, offering a structured basis for timing decisions grounded in WM theory.

The literature also highlights trade-offs in proactivity design. For example, overly proactive agents may be perceived as intrusive, while reactive ones risk missing key support opportunities \cite{wu_exploring_2021, park_generative_2023}. By modeling memory state, we aim to strike a more nuanced balance between providing the benefit of intervention and the cost of disruption.

\section{Design Rationale}
% - Explain why we choose WM framework for timing
% - The timing of assistance is based on the current user and environment context. From a user-centered perspective, this includes: the user's working context (what the user knows and don't know), their focus and attention (what are they considering now), and their mental capacity / interruptability
% - The working memory framework from psychology and cognitive science is great in that it has constructs that we can operationalize to monitor all of those elements

\begin{figure*}[h]

\centering
\includegraphics[width=\textwidth]{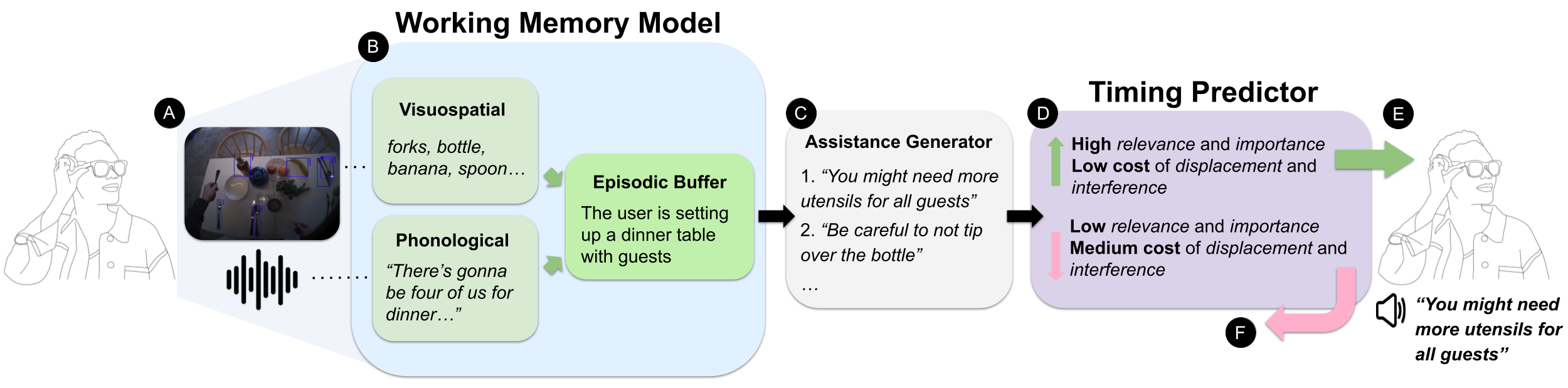}
%link to google drawing:
% https://docs.google.com/drawings/d/1WCgnyUbUPHLd2h_HV6nr9Z6nz8FcvOuZfmtc8amQkTw/edit
\caption{\textbf{\sys{} Workflow}. \textnormal{
\textcircled{\raisebox{-0.5pt}{A}}: Multimodal sensor input (visual and auditory) is captured from smart glasses. \textcircled{\raisebox{-0.5pt}{B}}: Perception memory encodes visual-spatial (e.g., “forks, bottle, banana”) and phonological (e.g., “There’s gonna be four of us for dinner”) signals, which are summarized into an episodic buffer describing the current task context. \textcircled{\raisebox{-0.5pt}{C}}: The assistance generator uses the working memory state to produce candidate messages. \textcircled{\raisebox{-0.5pt}{D}}: The timing predictor evaluates each message based on its predicted relevance and importance, as well as the cost of displacing existing memory or causing interference. \textcircled{\raisebox{-0.5pt}{E}}: Messages with high predicted utility are delivered as proactive voice assistance, while \textcircled{\raisebox{-0.5pt}{F}} lower-utility messages may be deferred for later evaluation or discarded.
}}
\label{fig:system-workflow}
\end{figure*}

To deliver timely support in dynamic, real-world tasks, proactive assistants must consider not just the external environment but also the user’s internal cognitive state \cite{amershi2019haiprinciple}. From a user-centered perspective, the optimal moment to provide assistance depends on multiple overlapping factors: (1) the user’s current \textit{mental context}—what they know, what they’re trying to do, and what might require support; (2) their \textit{attentional focus}—what modality or channel they are currently engaged with; and (3) their \textit{cognitive availability}—how much capacity remains to accommodate new input without inducing overload or disruption.

To capture these factors in a unified framework, we adopt a working memory (WM) model inspired by cognitive psychology and neuroscience \cite{baddeley_working_2012, cowan_magical_2010, miller_magical_1956}. Working memory is a lightweight and transient system responsible for maintaining and manipulating information over short time periods. It has several properties directly relevant to assistance timing—such as capacity limitations, decay over time, and competition between concurrent mental representations \cite{oreilly_making_2006, byrne_computational_1996}. Critically, WM reflects the user's immediate mental state—what they are actively attending to—making it particularly well-suited for predicting receptiveness to new information.

Unlike long-term user modeling or task-based heuristics, WM-based modeling focuses on what is mentally available right now. It allows the system to reason about real-time dynamics—what information is currently being held, how fresh or relevant it is, and whether new input would support or interfere with existing cognitive processes. Our model draws on classic tripartite WM theories that include visuospatial, phonological, and episodic components \cite{baddeley_working_2012, wastlund_experimental_2007}, and computational frameworks that simulate memory encoding, rehearsal, and interference.

This approach enables assistance to be grounded in the structure of ongoing mental activity—moving beyond surface behavior or static context to support proactive timing that is both informed and adaptive.

\subsection{Design Goals}
To implement this framework, we articulate three system-level goals that shaped the design of \sys{}:
\begin{itemize}
    \item \textbf{DG1: Facilitate Real-Time WM Modeling.} The system should continuously model the user’s working memory in near real-time to support fine-grained reasoning about when assistance is needed and how cognitively costly it might be. This includes maintaining up-to-date representations of memory items, their recency, modality, and relevance to ongoing context. This design goal stems from the temporal nature of WM, which decays over a short time and demands frequent updates \cite{reitman1974decaywithoutrehearsal, baddeley_working_2012}.
    
    \item \textbf{DG2: Utilize Observable and Multi-modal Inputs.} The WM model should be grounded in signals that are practically obtainable in everyday settings. We focus on visual and auditory information, aligning with the visuospatial and phonological stores of WM theory. By using wearable-friendly modalities such as video and audio, we ensure that the system approach remains deployable and applicable to other wearable devices.

    \item \textbf{DG3: Enable Cognitive-Informed Timing Decisions.} The WM model should support principled reasoning about when assistance should be delivered, deferred, or withheld. To do this, it should computationally process WM properties and mechanisms—such as memory encoding, decay, and interference —that reflect the user’s current mental state. These processes provide a structured basis for predicting the cognitive impact of potential interruptions and guiding assistance timing in a way that is both adaptive and grounded in WM theory.

    % \item \textbf{DG3: Simulate Computational Process for Timing Decisions.} The model should generate interpretable indicators that inform the value and cost of assistance at any moment. These include computational estimates of WM properties such as recency, interference, and relevance, which can then integrated into a utility function that predicts whether assistance should be delivered, delayed, or suppressed. This bridges cognitive theory with interaction design by enabling just-in-time support based on dynamic WM conditions.

    % \item \textbf{DG2: Observable and Multimodal Inputs.} The WM model should rely on signals that can be feasibly captured through wearable sensors, such as egocentric video and speech, enabling deployment in real-world settings without specialized instrumentation.
    
    % \item \textbf{DG3: Actionable Outputs for Timing Decisions.} The model should produce interpretable outputs—such as current WM state, available capacity, and potential interference—that can be used to reason about the value and cost of proactive assistance at any given moment.
\end{itemize}

\section{\sys{}}
\sys{} constructs a real-time WM representation using a pair of smart glasses with camera and microphone and computationally models memory mechanisms to predict the value and cost of potential proactive assistance.
Below we illustrate the system workflow with a user scenario walkthrough and detail the implementation.

\subsection{User Scenario Walkthrough}

Alex wears his smart glasses every day, relying on the onboard assistant to provide timely support for his everyday tasks. Today, Alex is setting up a dining table to welcome guests for dinner. As he moves through the room, the glasses passively capture his egocentric visual field and surrounding conversations through the built-in camera and microphone. Without any explicit input, \sys{} continuously models his real-time working memory state based on what he sees, hears, and interacts with (Fig.\ref{fig:system-workflow}.A).

As Alex begins by placing plates, cups, and cutlery on the table, the system encodes these visual signals into WM as memory items. When he places a coffee cup next to a spoon and hears someone say, \textit{“There’s gonna be four of us for dinner,”} \sys{} binds this input into a coherent episodic chunk labeled \textit{“The user is setting up a dinner table with guests.”} This episodic context is shown in Figure~\ref{fig:system-workflow}.B, where both visual (forks, bottle, banana...) and phonological (“\textit{There’s gonna be four of us for dinner…}”) inputs contribute to the episode.

With this chunk in WM and the table still in progress, the assistant proactively generates a candidate message: \textit{“You might need more utensils for all guests”} (Fig.\ref{fig:system-workflow}.C). Recognizing the message as highly relevant, important, and unlikely to interfere with the user's curent WM state, \sys{} immediately delivers the reminder (Fig.\ref{fig:system-workflow}.D,E). 
% The message about the bottle is less relevant to the dining setup and less important, so \sys{} defers this assistance for a later re-evaluation. 
The reminder to bring enough utensils arrives just as Alex is transitioning from cups to silverware, nudging him to adjust his setup without disrupting his focus.

Later, Alex notices a bottle of wine and places it near the edge of the table. \sys{} detects the placement and generates a potential reminder: \textit{“Be careful not to tip over the bottle”} (Fig.\ref{fig:system-workflow}.C).
However, before the system delivers the message, Alex's friend asks him how many eggs are left in the fridge.
As Alex goes to check, \sys{} transcribes the conversation and encode the new visual information of objects in the fridge, formulating a new episodic buffer representing this task context.
Withholding the comment about the wine bottle, \sys{}'s timing predictor identifies that Alex's WM is heavily engaged and that an interruption could cause confusion or disrupt task flow (Fig.\ref{fig:system-workflow}.D). 
The system defers the message and continues monitoring his cognitive state (Fig.\ref{fig:system-workflow}.F). Only once Alex completes egg-counting task and return to the dining table does \sys{} re-evaluate and determine that the cost of interference is low and the message still carries value. 
It then delivers the reminder, prompting Alex to nudge the wine bottle further from the edge just before guests arrive.
% Later, Alex notices a fruit bowl containing apples and bananas but hesitates about where to place them. Before the system can respond, a guest starts chatting about living room furniture. This shift in conversation causes Alex’s phonological WM to be dominated by information in another context. Though \sys{} has already generated a relevant reminder—“Put the apples aside for the fruit salad”—it predicts a high cost of interrupting the current memory context and defers delivery. Once the conversation subsides and the WM interference decays, the system re-evaluates and delivers the message at a more cognitively opportune moment.

As Alex steps back to review the table setup, \sys{} notices that no memory items reference napkins. Observing the WM content with lower task importance, it delivers one final reminder: \textit{“Don’t forget to put some napkins for your guests.”} The prompt lands at just the right moment—before Alex moves on from the current task context.

This scenario illustrates how \sys{} adapts its assistance timing based on moment-to-moment shifts in working memory. Rather than relying on fixed rules or scripted sequences, the system reasons over the dynamic structure of human cognition—delivering support that is timely, relevant, and less interruptive.

\subsection{Sensing and Information Encoding}
We build \sys{} on a pair of prototype smart glasses \cite{engel2023projectarianewtool}, equipped with RGB camera sensors and 7-channel audio microphone. This setup is in-line with common wearable device capabilities and modalities to increase extendability of our approach.

The smart glasses streams visual and auditory information to the system, hosted on a computer, as raw RGB images and audio buffer. We then utilize object-detection \cite{yolo11_ultralytics} and speech-to-text \cite{radford2022robustspeechrecognitionlargescale} models to extract visuospatial and phonological features in the user's environment.

To represent these information in the working memory model and enable similarity comparison, we embed the visual information (i.e. the visual image of the detected object and its text label) and the auditory information (i.e. the transcribed text of spoken speech) onto CLIP (Contrastive Language-Image Pre-Training) \cite{radford2021clip} and obtain vector representations of the detected information. 
We adopt CLIP for its versatility to encode both image and text information and facilitate semantic similarity comparison using the derived vector representations.

\begin{figure*}[h]

\centering
\includegraphics[width=\textwidth]{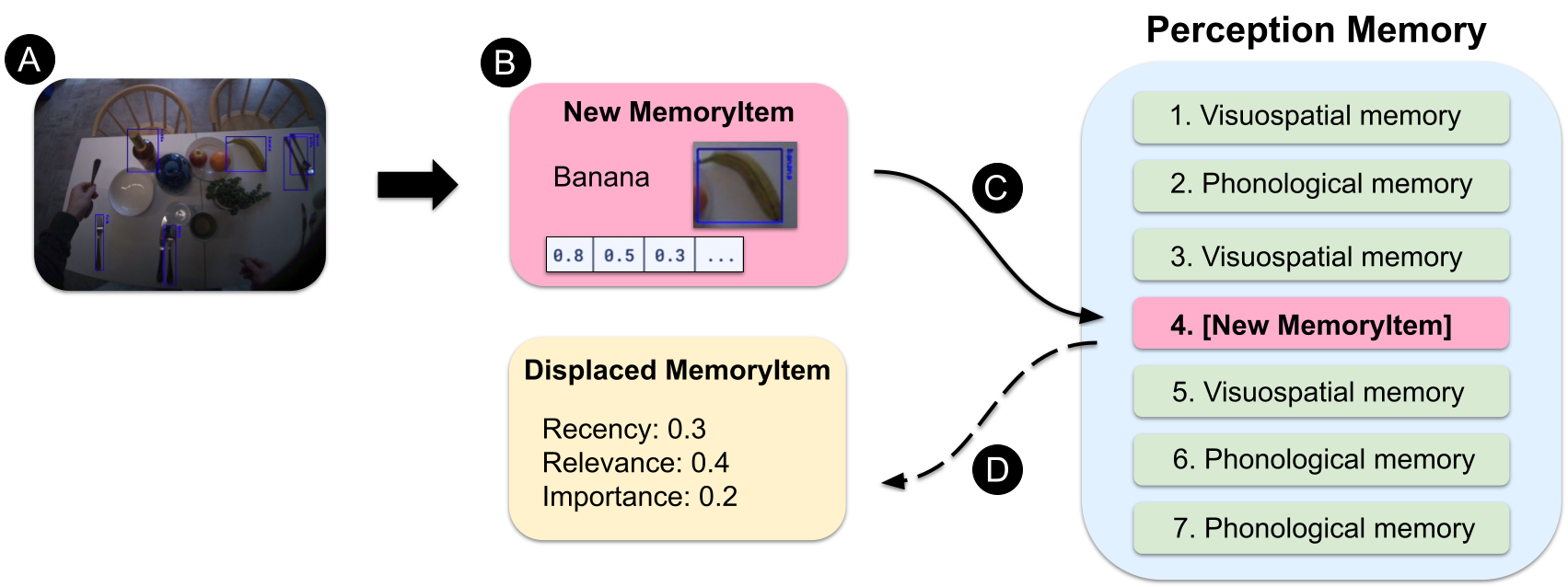}
%link to google drawing:
% https://docs.google.com/drawings/d/1BhmcsBUp_JSazcIMnn1acwtODTyAIWqSDbNyqs_Icp4/edit
\caption{\textbf{Working Memory Displacement}. \textnormal{As the user interacts with objects in the environment, 
\textcircled{\raisebox{-0.5pt}{A}} the system captures egocentric visual input as the user interacts with objects in the environment.
\textcircled{\raisebox{-0.5pt}{B}} A new memory item (e.g., a banana) is encoded with a corresponding feature embedding. If the perception memory is already at capacity, the system calculates the recency, relevance, and importance scores for existing items.
\textcircled{\raisebox{-0.5pt}{C}} The memory item with the lowest composite score (here: recency = 0.3, relevance = 0.4, importance = 0.2) is selected for displacement.
\textcircled{\raisebox{-0.5pt}{D}} The new memory item replaces the displaced item and is inserted into the updated perception memory store.
This mechanism ensures that the perception memory maintains the most updated information for supporting proactive assistance decisions.
}}
\label{fig:wm-displacement}
\end{figure*}

\subsection{Working Memory Model}
We are inspired by classic works in cognitive psychology to create a computational model of the user's working memory.
The most widely studied model proposed by Baddeley et al. theorized the working memory to contain four major components: a high-level central executive that manages and coordinates three lower-level components that store visual-spatial, phonological, and episodic information \cite{baddeley_working_2012}. 
We largely adopted this structure in \sys{} (Fig. \ref{fig:system-workflow}.B). In our system, the WM model is represented by two components: a low-level perception memory, which encodes visual-spatial and phonological information, and a high-level episodic buffer that distills perception memory content and formulates summarized descriptions about the user's current context. 
Different from Baddeley et al.'s model, the episodic buffer is derived from perception memory in our representation as internal episodes are not inherently observable.
This design allows us to leverage processed and structured sensor signals and use them to reason about plausible episodic buffer state.
% \mk{Might need to add justification on how our model differs from original}

Grounded by past studies and experiments, the perception memory has the capacity to store seven memory items \cite{miller_magical_1956} including both visuospatial and phonological information. We create a \texttt{MemoryItem} construct with attributes of a timestamp (time of encoding or last activation), type (visuospatial or phonological), content (text label and serailized image for detected object for visuospatial memory, transcribed text for phonological memory), and the feature vectors extracted from the CLIP embedding.

For the episodic buffer, Cowan et al. theorized that working memory content could be bounded and integrated into chunks of information, and that the working memory limit is four chunks of episodes \cite{cowan_magical_2010}. 
% Therefore, the episodic buffer component is operationalized as a list of four memory chunks \cite{cowan_magical_2010}.
Binding involves integrating features from different sources into an episode: \revise{an integrated chunk that} holds multidimensional information\cite{baddeley_working_2012}. Therefore, \sys{} represents chunks in the WM as a list of four custom type \texttt{MemoryChunk} objects, each containing a timestamp and the list of \texttt{MemoryItem} objects that are bound to this chunk. 

% Additionally, we can represent the \texttt{MemoryItem} objects as vectors on an embedding. To operationalize binding, we use an LLM to triage whether a new \textt{MemoryItem} should bind to an existing chunk, or we use metric-based calculations illustrated below.

\subsection{Memory Properties and WM Update}

To support real-time modeling of a user’s cognitive state, \sys{} continuously updates a structured representation of WM through mechanisms theorized by cognitive psychology. To facilitate the mechanisms, each \texttt{MemoryItem} is associated with three computationally defined properties—\textit{recency}, \textit{relevance}, and \textit{importance}—that reflect its current cognitive salience and value. These properties enable the system to simulate memory decay, identify key information, and make decisions about displacement and chunking as new information is processed.

\subsubsection{Memory Properties}

\textbf{Recency} captures how recently a memory item was encoded or rehearsed (i.e. reactivated to salience). Consistent with studies of short-term memory decay~\cite{berman2009verbaldecay, reitman1974decaywithoutrehearsal, mercer2014decaynonverbal}, we model recency as a linear function:
\[
\text{Recency} = 1 - \frac{t}{T}
\]
where $t$ is the elapsed time since encoding or last rehearsal, and $T$ is the maximum temporal threshold for short-term memory retention. Empirical studies suggest that items in working memory are actively maintained for approximately 15–30 seconds without rehearsal~\cite{Peterson1959ShorttermRO, reitman1971mechanismsforgetting}; we adopt $T = 30$ seconds as a reasonable bound in our system.

\textbf{Relevance} represents how semantically connected a memory item is to the user’s current WM context. We operationalize this as the average cosine similarity between the item's embedding and each episodic buffer summary. This formulation assumes that episodic chunks encode task-level goals or situations, making them a meaningful reference point for evaluating whether new information is on-topic or contextually grounded.

\textbf{Importance} measures the intrinsic value of the memory content, independent of context. It is assessed via an LLM prompt \ref{appendix:prompts-importance} that returns a score from 0 to 1. For instance, a memory item encoding \textit{“a fire alarm is going off”} may receive a high importance score (e.g., 0.9), whereas a commercial advertisement heard in the background may receive a low score (e.g., 0.1). This dimension helps prioritize information that may require urgent attention, even if it is unrelated to the current task.

These three properties are recomputed during each WM update and inform decisions about displacement, binding, and assistance timing.

\subsubsection{Memory Encoding and Displacement}
\label{section:Implementation-Encoding-and-Displacement}
New memory items are encoded from multi-modal sensor data and added to the perception memory, which has a fixed capacity of seven items~\cite{miller_magical_1956}. If there is space, the new item is inserted directly. Otherwise, an existing item will be displaced, simulating the overwriting behavior observed in human working memory under load~\cite{baddeley_working_2012,woodman2010displacevisual,luck2013visualcapacity} (Fig.\ref{fig:wm-displacement}).
\revise{To prevent repeated encoding of the same detected objects, the system compares each detected object to existing WM items using CLIP. If the detected objects match existing WM items (similarity \textgreater 0.95), the system deems the objects as already present in WM and updates item timestamps rather than encoding duplicates. }

To identify the item most likely to be displaced, we calculate a composite score for each memory item based on its recency, relevance, and importance:
\[
\text{Score} = \alpha \cdot \text{Recency} + \beta \cdot \text{Relevance} + \gamma \cdot \text{Importance}
\label{algorithm:composite-score}
\]
with default weights $\alpha = 0.3$, $\beta = 0.4$, and $\gamma = 0.3$ based on initial testing. The item with the lowest score is removed to make room for the newly encoded item (Fig.\ref{fig:wm-displacement}.D).

This mechanism allows the system to simulate both passive memory decay and displacement based on utility, providing a cognitively plausible and actionable model of memory dynamics.

\subsubsection{Memory Binding and Episodic Chunking}

After encoding, the system determines whether the new memory item should be bound to an existing episodic chunk in the WM’s episodic buffer. This reflects the psychological process of chunking, where related information is grouped into structured episodes for more efficient mental representation~\cite{cowan_magical_2010, thalmann2019chunking}.

Each chunk maintains a short textual summary generated by an LLM, along with the memory items it contains. To determine binding suitability, we compute a weighted score using two similarity metrics:
\begin{enumerate}
    \item \textbf{Episode Similarity:} Cosine similarity between the new item’s embedding and the episode summary embedding.
    \item \textbf{Item Similarity:} Average similarity between the new memory item and the current items within the chunk.
\end{enumerate}
\[
\text{Binding Score} = \lambda \cdot \text{Episode Similarity} + (1 - \lambda) \cdot \text{Item Similarity}
\]
We use $\lambda = 0.6$ to prioritize the task-level coherence captured in the episode summaries. If the highest score exceeds a default threshold $\theta = 0.5$, the item is bound to that chunk. Otherwise, a new chunk is created to represent this new memory item and a new episodic summary is generated using an LLM (Appendix \ref{appendix:prompts-episode}).

If the episodic buffer has already reached its capacity of four chunks~\cite{cowan_magical_2010}, the system displaces the least relevant chunk using the average composite value of its memory items. This maintains cognitive plausibility while allowing for dynamic restructuring as the user’s task evolves.

Together, these mechanisms enable \sys{} to simulate core WM behaviors—encoding, decay, displacement, and chunking—based entirely on real-time perceptual signals. These updates provide the foundation for reasoning about mental availability in the proactive timing model described next.

\subsection{Assistance Timing Prediction}
When \sys{} considers delivering proactive assistance, it invokes its timing predictor module (Fig.~\ref{fig:system-workflow}.D) to decide whether to issue the message immediately, delay it, or suppress it entirely. This decision is framed as a multi-objective optimization problem: the system seeks to balance the potential benefits of assistance against its cognitive costs. Specifically, \sys{} aims to:

\begin{itemize}
    \item Maximize the \textbf{value of the assistance}—how beneficial the new information is to the user given their current mental context.
    \item Minimize the \textbf{cost of the interruption}—how disruptive the intervention might be to the user’s ongoing cognitive processing.
\end{itemize}

We model this tradeoff using the following utility function:
\[
\text{Utility} = (W_I \cdot I + W_R \cdot R) - (C_D + C_I)
\]
where $I$ (Importance) and $R$ (Relevance) characterize the value of the proposed assistance, $C_D$ is the predicted cost of displacing memory content, $C_I$ is the predicted interference cost, and $W_I = 0.6$, $W_R = 0.4$ are tunable weights. The system treats each candidate assistance message as if it was a new \texttt{MemoryItem} to be encoded into perception memory, and evaluates its impact accordingly. We describe computational details to each parameter below.

\subsubsection{Maximize Assistance Value}

The value of delivering the assistance is predicted by the value of encoding the new information to the user's current WM state. We quantify the value of assistance by calculating two memory properties: Importance ($I$) and Relevance ($R$). 

Before an assistance is delivered, \sys{} evaluates the Importance and Relevance scores of the assistance using the same computational approach described above (Section \ref{section:Implementation-Encoding-and-Displacement}). We do not consider Recency as a key factor in maximizing the value as if the assistance is delivered, the recency of the encoded information would already be maximized. 

Since Importance and Relevance are calculated by themselves, each has a range of [\textit{0,1}], making the predicted value of assistance to have a range of [\textit{0,2}].

\begin{figure}[h]

\centering
\includegraphics[width=\columnwidth]{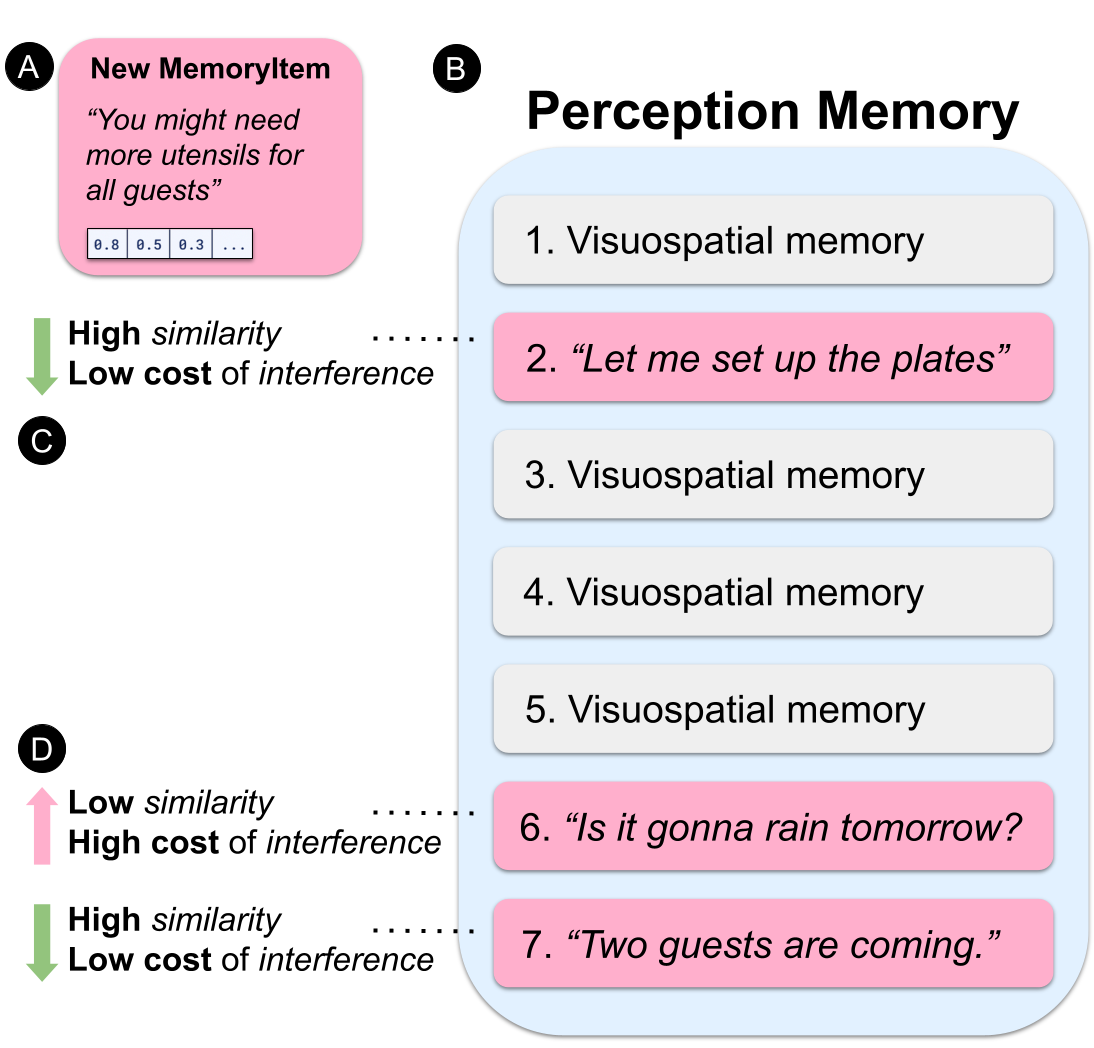}
%link to google drawing:
% https://docs.google.com/drawings/d/1jtStTEjF1_jdlYfaEHvq9zRf7LPf-yBscVTwh70ILrc/edit
\caption{\textbf{Working Memory Interference}. \textnormal{Interference is computed for each memory item in the same modality (e.g., visual or auditory) using cosine similarity.
\textcircled{\raisebox{-0.5pt}{A}} A new proactive assistance message (e.g., \textit{“You might need more utensils for all guests”}) is evaluated for delivery.
\textcircled{\raisebox{-0.5pt}{B}} \sys{} identifies memory items in the perception memory that share the same modality—in this case, phonological.
\textcircled{\raisebox{-0.5pt}{C}} For each overlapping memory item, interference is computed using the formula $(1 - \text{cosine similarity})$ based on CLIP embeddings. Highly similar memory items can be chunked and integrated, thus leading to lower interference costs.
\textcircled{\raisebox{-0.5pt}{D}} Conversely, dissimilar memory items cause higher interference to maintain. The interference costs are aggregated and normalized to [\textit{0,1}] to estimate how disruptive the new message would be. If cost is high, delivery may be deferred.
}}
\label{fig:wm-interference}
\end{figure}

\subsubsection{Minimize Interruption Cost}

We model the cost of interruption based on two main factors: displacement ($C_D$) and interference ($C_I$).
The cost of displacement is calculated similar to the displacement process during \texttt{MemoryItem} encoding (Section \ref{section:Implementation-Encoding-and-Displacement}).
We predict the potential information loss on the user's WM content if an assistant introduced new information to the user.
Framing the displacement as a cost to be minimized discourages delivering assistance when it would overwrite highly valuable information.
\revise{For this cost of the predicted displacement ($C_D$), we calculate a composite score for each existing WM item and find the minimum, representing the item to be displaced.}
The composite score is calculated using the same formula as in Section \ref{algorithm:composite-score} with weighted Recency, Relevance, and Importance and has a value range of [\textit{0,1}].

On the other hand, the cost of interference reflects the attentional disruption caused by an incoming message when it uses the same modality (e.g., auditory) as currently active memory items (Fig.\ref{fig:wm-interference}). 
We model the cost of interference on both the \textit{modality overlap} between the assistance information and the existing WM content and the \textit{semantic integration potential}.
Following Baddeley's model \cite{baddeley_working_2012}, phonological interruptions (e.g., voice messages, conversations) compete with verbal WM content, while visual interruptions affect visuospatial WM.
\sys{} leverages the \textit{episodic buffer} mechanism \cite{baddeley_working_2012, cowan_magical_2010} to enable chunking of schema-congruent information (i.e. information that are similar). 
When assistance semantically aligns with existing WM content, it can be integrated rather than competing, reducing effective interference \cite{van2012schema,van2020congruency}.
In contrast, introducing dissimilar information with modality overlap induces higher interference, as it requires more effort to maintain existing WM state.
While feature-overlap theories \cite{oberauer2009interference} suggest similarity between memory information could instead increase interference in recall and recognition tasks, our system operates in a physical environment where users are not actively memorizing and recalling information, but rather encoding system assistance to aid their current task.

\revise{Figure \ref{fig:wm-interference} illustrates an example of WM interference. The candidate proactive assistance voice message from ProMemAssist—“You might need more utensils for all guests”—is represented as a new phonological memory item (A). \sys{} compares this item to existing phonological items in the perception memory (B), which currently contains a mix of visuospatial and phonological memory items. Only items in the same modality are considered for interference. Among the phonological items, “\textit{Let me set up the plates}” (2) and “\textit{Two guests are coming}” (7) are semantically related to the new message under the overall task context of setting up a dinner table, resulting in high similarity and low interference (C). In contrast, “\textit{Is it gonna rain tomorrow?}” (6) is less relevant, leading to low similarity and high interference (D).}
% The overall interference cost is calculated by aggregating similarity scores and normalized to [0,1] to inform whether the message should be delivered. Lower interference suggests the message can be more easily integrated into working memory without disruption.

We compute the raw interference cost based on semantic dissimilarity between the candidate assistance and existing memory items in the same modality (e.g., visual or auditory):
\[
C_I' = \sum_{m \in \text{WM}_{\text{SameModality}}} (1 - \text{Similarity}(m, \text{Assistance}))
\]
Here, similarity is measured using cosine similarity of CLIP embeddings.
To ensure that $C_I$ is normalized to the range [\textit{0,1}] like the other terms in the utility function, we divide the sum by the number of comparisons:
\[
C_I = \frac{C_I'}{|\text{WM}_{\text{SameModality}}|}
\]
This yields a normalized interference score that penalizes semantically incongruent interruptions more severely, particularly when they overlap with active memory channels. The overall cost of interruption (displacement and interference), then, is confined by the same range as the value of assistance: [\textit{0,2}].

\subsubsection{Timing Decision Rule}

After computing the utility score of a candidate assistance, \sys{} applies a threshold-based policy to decide whether, when, or if the message should be delivered. If the utility score exceeds a predefined threshold ($0.75$ by default from testing), the message is delivered immediately, indicating that it is contextually relevant, important, and unlikely to be disruptive.

If the utility score is greater than 0 and below the threshold, the message is held in a deferred queue. These deferred messages are re-evaluated on subsequent WM updates to determine whether changing cognitive conditions (e.g., reduced interference or increased relevance) make them more suitable for delivery.
Messages with a utility score less than or equal to 0 are discarded, as they are deemed unlikely to provide meaningful benefit or would impose too high a cognitive cost.

This staged decision strategy allows \sys{} to reason not only about the appropriateness of intervention at a given moment, but also to revisit borderline cases as the user’s mental state evolves—producing more thoughtful, context-sensitive assistance over time.

\subsection{Proactive Assistance Generation}
The primary contribution of \sys{} lies in modeling working memory to inform the timing of proactive assistance. Our goal is not to generate optimal or goal-directed assistance content, but to explore when such assistance should be delivered based on the user’s cognitive state. However, to support evaluation of different timing strategies, we implement a lightweight assistance generation module to simulate plausible messages a smart glasses assistant might produce (Fig.\ref{fig:system-workflow}.C).

This assistance generation module is triggered at every working memory (WM) update. It uses an LLM to produce a candidate voice message grounded in the current cognitive context. The prompt (Appendix \ref{appendix:prompts-sys-generate-assistance}) to the LLM includes:
\begin{itemize}
    \item The most recent \texttt{MemoryItem} added to perception memory (e.g., an object seen or speech heard)
    \item A summary of episodic buffer chunks representing the user’s recent context
    \item A short history of prior generated and delivered assistance messages
\end{itemize}

The LLM returns candidate assistance messages with Importance scores, if it deems necessary after evaluating the user's WM state and context. These messages are then passed to the timing predictor module, which evaluates whether, when, and how they should be delivered based on their predicted cognitive impact. If the LLM does not think any assistance is required, it returns no messages.

\begin{figure*}[h!]

\centering
\includegraphics[width=\textwidth]{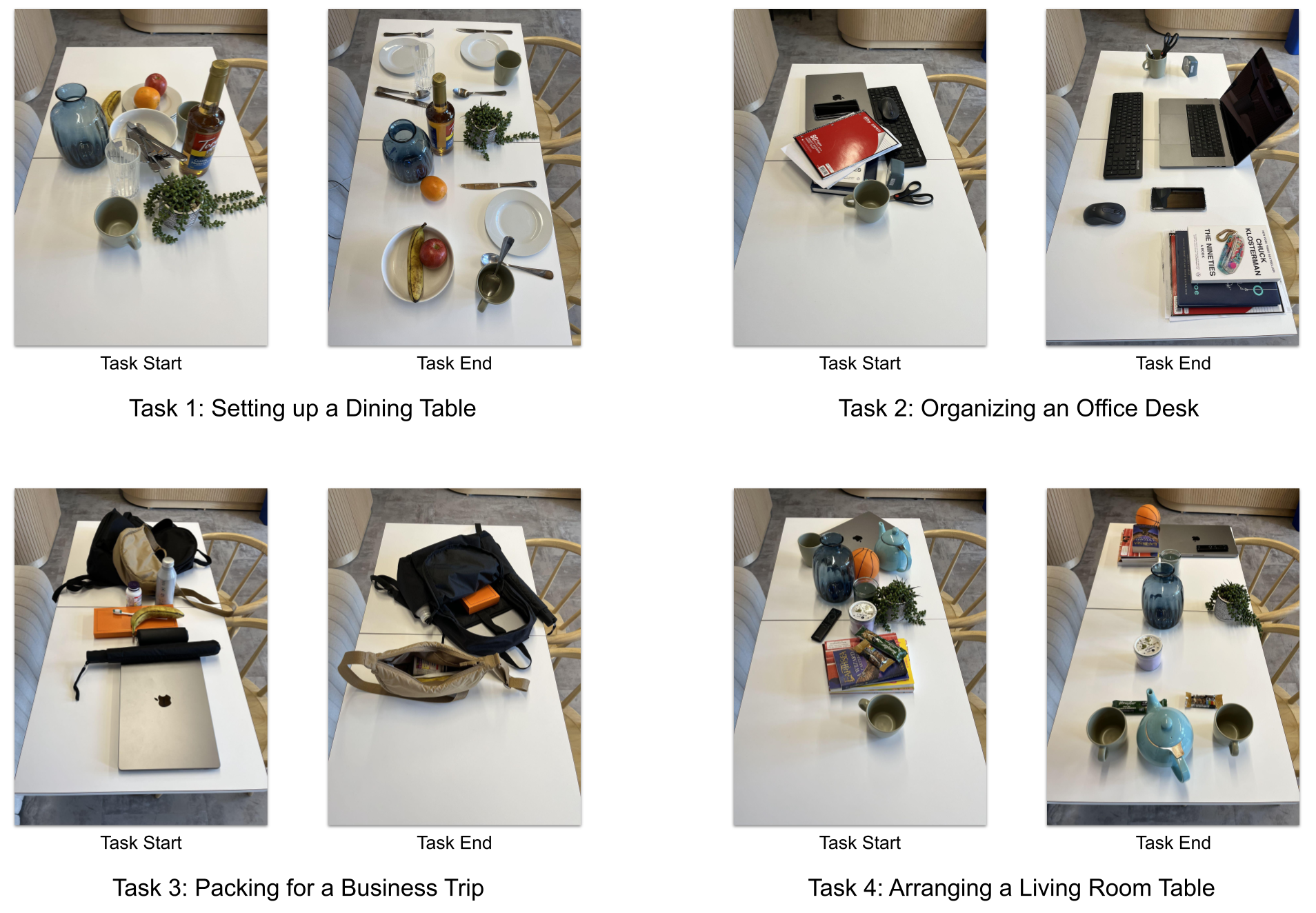}
%link to google drawing:
% https://docs.google.com/drawings/d/1DcZgs3iRFx_dM9XdLW5_YnaLBGqIvMjqGgGztnow8o0/edit
\caption{\textbf{Task Settings}. \textnormal{Four task scenarios were used in the user study: Setting up a dining table, Organizing an office desk, Packing for a work trip, and Arranging a living room table. Each task has a fixed starting position of objects (task start), and participants are asked to organize, arrange, and sort objects in the way they see fit, resulting in finishing positions like illustrated in the task end images. }}
\label{fig:task-setting}
\end{figure*}

\subsection{System Implementation}

\sys{} is implemented on a research prototype smart glasses equipped with an RGB camera and a 7-channel microphone array~\cite{engel2023projectarianewtool}. We use the glasses to stream raw video and audio at 1-second intervals to an off-the-shelf laptop computer, where feature extraction is performed using object detection (YOLOv11) \cite{yolo11_ultralytics} and speech-to-text models (Whisper) \cite{radford2022robustspeechrecognitionlargescale}. Visuospatial and phonological information is embedded via CLIP~\cite{radford2021clip}, and stored in structured WM representations.

Memory management, binding, timing prediction, and LLM prompting run on a lightweight Python pipeline on the companion laptop computer. The system-generated assistance are delivered in the form of voice messages on the companion device, as the smart glasses prototype does not have speakers. All components operate in near real-time, with updates processed incrementally at each input interval of one second.

\section{Evaluation}

To evaluate the effectiveness of \sys{} in delivering timely and non-disruptive proactive assistance, we conducted a within-subject user study comparing the WM-modeling timing strategies in our system against a baseline LLM-based system without WM constructs. Our primary research question was whether modeling working memory improves the perceived timing and appropriateness of proactive assistance, particularly in cognitively demanding, real-world tasks.

\subsection{Participants}

We recruited 12 participants (8 male, 3 female, 1 non-binary, mean age 36 y.o.) from an internal participant pool. All participants were research engineers or scientists familiar with consumer wearable and mixed reality devices, though none had prior experience with the study system. Sessions lasted approximately 60 minutes per participant.

\subsection{Study Design}

We used a within-subject design with two conditions:
\begin{itemize}
    \item \textbf{\sys{}}: Assistance timing was governed by our WM-based timing predictor, which continuously evaluated assistance value and interruption cost.
    \item \textbf{Baseline}: Assistance timing was governed directly by a prompt-engineered LLM, which was given access to the same multi-modal observations and task context but did not include any explicit memory modeling. In this condition, the LLM was responsible for both generating assistance content and deciding whether or not to deliver it based on a reasoning prompt.
\end{itemize}

We designed the Baseline system to represent a strong and realistic comparator: a generative model with heuristic reasoning capabilities but without WM-state awareness. The Baseline LLM was instructed to simulate timing sensitivity through prompt-based guidance. Its system prompt emphasized relevance, urgency, and non-redundancy of assistance, and included principles for withholding low-value or interruptive messages (Appendix~\ref{appendix:prompts-baseline-generate-assistance}).

Unlike \sys{}, which separates timing decisions from assistance generation and reasons over a dynamic memory model, the Baseline condition relies on the LLM's ability to perform implicit cost-benefit reasoning within a single-turn prompt. This design simulates how exisiting proactive assistants might operate—leveraging LLMs to infer user state and provide helpful nudges based solely on task context and heuristics, without access to internal cognitive structure.

By holding perceptual inputs, task scenarios, and generative capabilities constant across both conditions, our study isolates working memory modeling as the key factor for evaluating differences in timing, user experience, and perceived assistance quality.

\subsection{Tasks}

To evaluate the system’s ability to support working memory–informed assistance timing, we designed four tabletop tasks that simulate common daily activities involving both hands-on object interaction and contextual reasoning. Each task required participants to physically manipulate everyday items while tracking conversational cues, short-term goals, and shifting priorities.

The four scenarios (Fig.\ref{fig:task-setting})—\textit{setting up a dining table}, \textit{organizing an office desk}, \textit{packing for a trip}, and \textit{styling a living room table}—were intentionally chosen to reflect different spatial configurations, object types, and social contexts. 
Participants were asked to place, pack, or group objects as they saw fit, while the experimenter interjected with scripted relevant or irrelevant information to simulate a cognitively active and distraction-prone environment.
For example, while setting up the dining table, a relevant prompt might be \textit{“The spoon is for the coffee in case anyone needs stirring,”} which encourages the participant to consider object-function alignment. An irrelevant prompt might be \textit{“I'm thinking of redecorating the living room, what do you think about a new couch?”} which is contextually unrelated to the current task. 
These scripted interjections were delivered either in response to the participant interacting with relevant objects (to introduce decision-making pressure), or during natural pauses and transitional moments in the task, when cognitive load was likely to be lower. 
% This approach follows prior methods in multitasking and interruption research \mk{cite}, where interruption timing is varied to reflect realistic fluctuations in attention and workload.

% These tasks were designed to elicit active working memory engagement. They required participants to:
% \begin{itemize}
%     \item Maintain and update object-related goals in response to environmental and conversational changes;
%     \item Shift attention between physical tasks and social interactions;
%     \item Retain context across multiple modalities (visual, spatial, verbal);
%     \item Manage competing demands and potentially discard or reprioritize information.
% \end{itemize}

We selected this design to create a realistic setting where mental load naturally fluctuates and timely assistance becomes both necessary and risky, thus providing a strong testbed for evaluating proactive timing strategies. 
See Appendix~\ref{appendix:task-details} for full details on task setup and experimenter prompts.

% \subsection{Tasks}

% We designed four tabletop tasks intended to simulate common daily activities that require both physical interaction and mental tracking of multiple items or goals:
% \begin{itemize}
%     \item \textbf{Setting up a dining table}: Participants arranged dinnerware based on a visual and verbal reference.
%     \item \textbf{Organizing an office desk}: Participants sorted and grouped office supplies to prepare a workspace.
%     \item \textbf{Packing for a work trip}: Participants packed items into a bag according to spoken instructions and contextual hints.
%     \item \textbf{Arranging a living room table}: Participants styled a coffee table using decorative items and real-time audio suggestions.
% \end{itemize}

% These tasks were selected to engage both the visual-spatial and phonological components of working memory, while allowing for observable interaction with physical objects. Each task required sustained attention, working memory updates, and occasional multitasking. To simulate a realistic cognitive load, each task was paired with scripted conversational prompts from the experimenter to maintain a dynamic, distraction-prone environment.

% This design ensured that participants would naturally accumulate and refresh WM content throughout the task, allowing us to evaluate the system’s ability to reason about timing in cognitively active scenarios.

\subsection{Procedure}

Participants were introduced to the study as a test of timing strategies for smart glasses assistants.
\revise{They were instructed to wear smart glasses, perform physical tasks, and receive AI assistance and occasional interaction from the experimenter. No information about system mechanism or conditions was disclosed until after the interview and debrief.}
Each participant completed two tasks under each condition, with system condition order counterbalanced to minimize learning effects.
After each task, they filled out a survey evaluating the timeliness, sense of interruption, relevance, and helpfulness of the system’s assistance, as well as task load via NASA-TLX\cite{hart1988nasatlx}.
After completing all four tasks, participants took part in a semi-structured interview, where they were asked to reflect on the system's behavior, particularly moments when it helped or disrupted their workflow, how they felt about the timing of the proactive assistance, their sense of control, agency, and trust towards the system.
\revise{The full study procedure can be found in Appendix\ref{appendix:study-procedure}.}

\begin{figure*}[h]

\centering
\includegraphics[width=\textwidth]{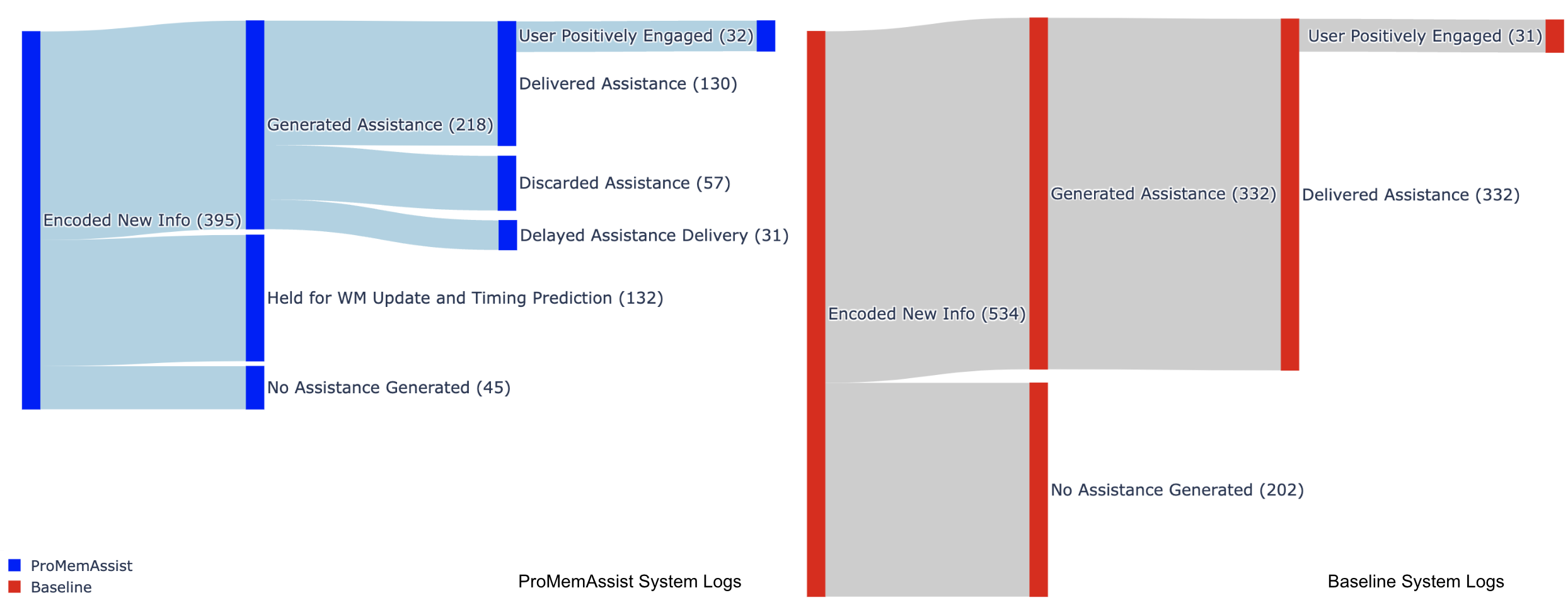}
% new link to google drawing:
% https://docs.google.com/drawings/d/1qBpApJ1SXbFJNC093ST5d441OQ2NFpW-X_zse20Lxew/edit
\caption{\textbf{Overall system behavior from ProMemAssist (left) and the Baseline (right).} \textnormal{ProMemAssist selectively filters and delays assistance based on WM state, resulting in fewer overall messages but a higher proportion of positive user engagement. The baseline uses an LLM to evaluate whether assistance is necessary to be generated and delivers all messages at time of generation.}}
\label{fig:system-log}
\end{figure*}

\subsection{Data Analysis}

For each session, we collected system-level logs to record the behavior and internal state of both \sys{} and the baseline system. These logs included timestamps and metadata for each assistance message generated, delivered, deferred, or discarded. Participants' positive engagements with assistance were manually logged by the experimenter in real-time using keypresses on the companion laptop device. Positive engagements were defined as when participants visibly or verbally responded—for example, by acknowledging the message or acting on the suggested action. These annotations were later validated against video recordings. 
% We focused on positive engagement and did not label cases where participants ignored or dismissed the assistance.

Post-task surveys were used to assess participants’ perceptions of assistance timing, helpfulness, and relevance, as well as subjective workload using the NASA-TLX (noted as Q5 to Q10 in Fig.\ref{fig:survey}). 
\revise{We adapted the raw NASA-TLX scale to a 1-7 Likert scale to reduce decision fatigue over four rounds of ratings. The 7-point scale was chosen to maintain scale validity and discriminability \cite{Preston2000OptimalNO,Lewis2017UserER}.} 
To analyze the data, we used paired $t$-tests to compare continuous measures between conditions and Wilcoxon signed-rank tests for Likert-scale survey data.

We also conducted a thematic analysis of the semi-structured interview transcripts. In addition to open-ended reflections on system behavior, participants were asked what factors an intelligent assistant should consider when deciding when to intervene. 
Their responses were coded to identify recurring themes, such as mental availability, task completion stages, and social or contextual cues. 
These qualitative insights complemented our quantitative findings and informed the broader design implications of WM-based timing.

\section{Results}
\subsection{\sys{} Delivered More Selective Assistance And Received More Positive Engagement}

We first analyze the two systems' behavior to generate and deliver assistance to users based on WM-based and LLM-prompt-based strategies. 
The system logs (Fig.\ref{fig:system-log}) revealed that \sys{} encoded new sensor information and processed it in the WM model 395 times. Out of those occurrences, \sys{} generated 218 (55.2\%) candidate assistance messages, of which 130 were delivered immediately, 31 were deferred for future re-evaluation, and 57 were discarded due to low predicted utility. 
In contrast, the baseline system encoded new information 534 times. Using the prompt-engineered LLM (\ref{appendix:prompts-baseline-generate-assistance}), the baseline evaluated sensor information and deemed it necessary to generate and deliver assistance 332 times (62.2\%). Notably, \sys{} and the baseline system used similarly prompted LLMs with the same tuning to generate assistance at comparable rates (\sys{}: 55.2\%; Baseline: 62.2\%). However, \sys{}'s additional Timing Predictor selective delivered assistance based on the WM-based context modeling.
We did not identify a significant difference in task completion time across two conditions (\sys{}: $mean=175$ seconds; Baseline: $mean = 182$ seconds).
\revise{Additionally, we report the WM model’s performance in the ProMemAssist condition. On average per task, the system processed 16.5 encoding events. Of these, 6.92 resulted in new MemoryItems being added to the working memory, 4.67 were identified as repetitions of existing items (and thus not added to the WM), and 4.88 involved replacing an existing MemoryItem due to capacity constraints.}

Due to the different timing strategy, \sys{} received more positive user engagements, such as verbal confirmation of the timeliness or usefulness of the assistance (e.g. \textit{``I needed that info.''}), or direct follow-up action in response to the assistance (e.g. packing the medication after the proactive reminder). 
By coding the participant's reaction to proactive assistance, we found that participants responded positively to 32 (24.6\%) of the delivered messages in the \sys{} condition, higher than the 31 (9.34\%) positive responses to delivered messages in the baseline condition.
In the baseline condition, participants often ignored the proactive assistance due to interruption to current task focus which forces context-switching.
% P4 expressed that the assistance from baseline ``\textit{was just so irrelevant that I just... completely filtered it out of my memory...It's because the system is not aware of your state.}''
\revise{This is due to the fact that} the system did not model the mental state of the user. Although the LLM is instructed to provide high-value assistance and avoid interruption (the same as in \sys{}), this approach alone was not structured enough to time the proactive assistance well.
P5 remarked ``\textit{I felt like if I'm currently working on some task and then I have, like, some cognitive load, you shouldn't tell me too much, unless it's important.}''
These results indicate that WM-informed filtering led to more selective delivery of proactive assistance, improving the user engagement overall.

\subsection{ProMemAssist Better Aligned with Mental Availability and Task Flow}

We next analyzed participants’ qualitative reflections on how the two systems aligned with their mental availability and supported task flow. 
Some participants expressed that \sys{} delivered assistance at moments that felt less disruptive and more cognitively appropriate, often waiting until the user was mentally unoccupied or between subtasks.
For instance, P9 noted, “\textit{I liked that it [\sys{}] wasn’t always talking to me when I was in the middle of something. It felt like it waited until I was done.}”
Similarly, P3 highlighted that timing felt well-matched to their mental state, especially toward the end of a task: “\textit{[\sys{} intervened when] you're almost done and you don’t have as much on your mind — definitely yeah, [mental capacity] definitely matters.}”

Participants attributed the improved timing to \sys{}’s ability to recognize when they were cognitively engaged or overloaded. This suggests that even when participants couldn’t precisely describe how \sys{} worked, they implicitly recognized its sensitivity to their attentional bandwidth.

Moreover, participants frequently referenced the influence of broader factors—such as task stage, emotional readiness, and individual preference—on their receptiveness to assistance.
P6 explained, “\textit{There were moments where I wanted help and moments where I didn’t want anything. It kind of depends where you are in the task.}”
P12 echoed this idea, saying, “\textit{Depending on the mood that I'm in, I'm much more receptive to different levels of technological intervention.}”

These reflections reinforce the premise of \sys{}: proactive support should not be delivered uniformly based on predefined rules or context, but instead carefully timed based on the user’s mental state. Participants described a subtle but meaningful improvement in how assistance was delivered, aligning with their shifting cognitive states and lowering interruptions.

\begin{figure*}[h]
\centering
\includegraphics[width=\textwidth]{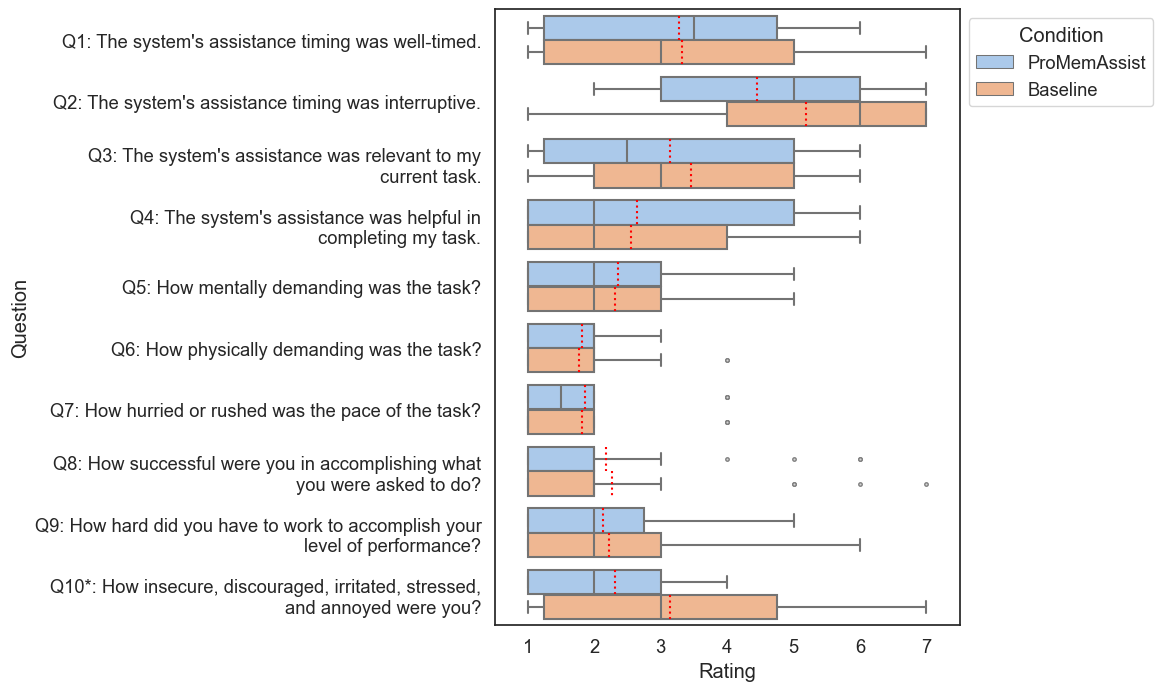}
\caption{\textbf{Likert-scale responses comparing \sys{} and baseline conditions.} \textnormal{Box-and-whisker plots for each question show participant ratings across the two conditions. The red dotted lines indicate mean values. Anchors ranged from 1 (Strongly disagree, Very low) to 7 (Strongly agree, Very high). \revise{Q5 to Q10 are questions adapted from NASA-TLX \cite{hart1988nasatlx}.} For Q8, 1 is Perfect and 7 is Failure \cite{hart1988nasatlx}.  Q10 showed a statistically significant difference ($p = 0.043$), suggesting lower perceived \revise{frustration} in the \sys{} condition.}}
\label{fig:survey}
\end{figure*}

\subsection{\sys{} Led to Less \revise{Frustration} and Less Perceived Interruptions}
% \subsection{\sys{} Reduced Stress and Felt Less Interruptive Despite Comparable Ratings on Helpfulness and Timing}

To evaluate perceived workload and system experience, we analyzed participant responses to post-task Likert-scale surveys (Fig.\ref{fig:survey}). Among all NASA-TLX and subjective metrics, \revise{frustration (Q10)} was the only dimension with a statistically significant difference: participants reported lower insecurity, discouragement, irritation, stress, and annoyance in the \sys{} condition (mean = 2.32) compared to the baseline (mean = 3.14), with $p < 0.05$ (Fig.~\ref{fig:survey}). This reduction in \revise{frustration} aligns with participants’ qualitative reports of smoother task flow and fewer disruptive moments.
P2 described “\textit{It felt like I was more in control [in \sys{}], even when it reminded me of things. It was helpful, not pushy.}” 

Other workload dimensions such as mental demand, physical demand, time pressure, and task difficulty showed no significant differences ($p > 0.5$)
We attribute this to the short, fast-paced nature of our tabletop tasks, which were intentionally designed to tax working memory over a constrained period. While this setup elicited high cognitive load, it may have limited participants’ ability to discern finer-grained differences in system support on workload measures.

Participants rated assistance from \sys{} as less interruptive (mean = 4.45) than baseline (mean = 5.18), although the difference was not statistically significant ($p = 0.158$).
While participants qualitatively reported a more interruptive baseline experience, they occasionally acknowledged its benefit of higher recall coverage due to delivering more frequent assistance. 
% As one participant explained: 
% INSERT QUOTE HERE
This suggests that while baseline’s high-frequency delivery increased the chance of timely reminders, it also led to more false positives and perceived disruption. \sys{}, in contrast, delivered fewer but more carefully timed messages—reflected in the significantly higher proportion of positively engaged responses (Section 6.1).

We also \revise{did not} observe a significant difference in ratings on whether the assistance was well-timed.
Interestingly, we found participants sometimes responded favorably to any assistance that happened to align with their task needs—regardless of whether the timing was optimal. 
We further discuss the coupling effect between proactive assistance quality and timing in the Discussion.
Additionally, given the rapid, multitasking nature of the study tasks, participants had limited time to reflect on and differentiate between more subtle nuances between degrees of timely assistance, whereas interruptions are more salient and memorable.

Importantly, we observed no significant differences in ratings of assistance relevance or helpfulness. This outcome is expected as in our experimental design, both \sys{} and the baseline system used the same underlying LLM to generate assistance content to ensure comparable message quality across conditions. This isolates \textit{timing of delivery} as the key experimental variable influencing user engagement and experience.

% Second, given the brief, high-tempo nature of the tasks to examine the effects of the working memory, rather than long-term memory, participants had limited opportunity to fully reflect on or contrast the subtle timing differences across conditions. 
% Still, descriptive statistics revealed small trends in favor of \sys{}, suggesting that with longer or more complex task contexts, the cognitive benefits of WM-based timing may become more salient.
Overall, these findings suggest that WM-based timing can improve user experience—especially by reducing \revise{frustration}—without requiring changes to the underlying assistance content. Future studies involving longer tasks, higher stakes, or multi-session use may better reveal the cognitive benefits of just-in-time support and its interaction with perceived message quality.

\subsection{Participant Desired Long-Term Personalization And Feedback Mechanism}

While participants generally appreciated the cognitively aware timing in \sys{}, many emphasized that effective assistance requires more than good timing—it also demands contextual understanding and personalization. Several participants pointed out instances where assistance was mistimed or misaligned with their current situation or goals.
P7 expressed \textit{“I think once it has some context about my activities… it works much better. Like where I stay, something like that... then it knows my question is regarding something like that.”}
This insight point to a broader challenge: even well-timed assistance can fall short if the system lacks a deeper model of user intent, environment, or long-term goals. While our current implementation focuses on modeling mental availability through WM dynamics, future extensions could integrate richer user modeling, such as long-term memory, goal tracking, or environment-aware grounding.

Participants also expressed a strong desire for adaptivity and feedback mechanisms—suggesting that intelligent assistants should not only reason about the user’s state but also learn from it.
Like P4 described \textit{“It should learn from me over time. Like, I always forget my keys, tell me that automatically.”}
P7 also felt \textit{“there's a lack of feedback to this... the feedback loop is kind of not there. It is helping me but it's like we are walking parallelly.”}
These reflections suggest promising directions for future research: developing systems that learn personalized timing preferences, accept corrections or confirmations from users, and refine their timing decisions over time. Feedback-aware systems could also better calibrate when to interrupt by learning when users tend to respond positively or ignore assistance.

Overall, these findings reinforce our design motivation: \sys{} takes an initial step toward more cognitively aligned proactive assistants by modeling working memory and attentional load. However, it also highlights the importance of viewing timing as part of a broader framework—one that includes content grounding, adaptivity, and co-adaptive interaction over time.

\subsection{\revise{Control and Trust Depended Less on Timing and More on Assistance Quality}}
\revise{From the interview, participants generally felt in control of their task performance, but their perceived control over the system’s behavior and their trust levels varied. Five participants reported high system control and agency, describing that they could freely choose whether or not to respond to assistance, regardless of whether the assistance was timing or interruptive. They felt that the system’s suggestions were non-intrusive in the physical task environment and allowed them to remain in charge of their actions. Conversely, five others felt a lack of system control, often pointing to the absence of a feedback mechanism. Without the ability to tune the frequency or the area of assistance, they felt the system operated independently of their input. Two participants articulated a nuanced view, describing low control over the system’s behavior, yet high agency in how they executed tasks. This suggests that the WM-based timing of assistance did not play a significant role in participants' perception of control, but system designs such as feedback channels to adjust system behavior could increase the sense of control.}

\revise{Trust in the system was also influenced more by the quality of assistance than by its timing. Participants were sensitive to perceived utility of the assistance. For instance, P5 noted that a single unhelpful or misleading suggestion could break trust and deter future usage of the system. These findings align with broader literature on proactive systems, where relevance and usefulness are critical to sustained user trust. They also underscore the importance of designing assistants that not only interrupt appropriately but provide consistently useful content, and that allow users to shape or calibrate interaction over time.}

\section{Discussion}

\subsection{Design Implications for Cognitive-Aware Assistants}
Our study demonstrates that modeling working memory in real time can support more thoughtful and less disruptive proactive assistance. By incorporating cognitive constructs such as recency, capacity limits, and modality-based interference, \sys{} was able to selectively deliver assistance in ways participants found better aligned with their mental state and task focus.

These findings offer key implications for the design of future always-on, wearable, and AR-based assistants. As these systems increasingly operate autonomously and continuously in users' daily lives, cognitive modeling can serve as a critical filter to minimize intrusiveness and support mental well-being. Rather than simply reacting to environment cues or predefined triggers, assistants should reason about the user's attentional state and mental load before intervening.

At the same time, these systems raise new questions about safety, privacy, and control. Always-on sensing—especially involving visual and audio data—can create concerns about data security, even when used locally. Designers of cognitive-aware systems must balance the need for rich real-time signals with user agency, providing transparent, privacy-preserving options for when and how memory modeling occurs.

Importantly, cognitive-aware assistants may need to err on the side of caution. Our results suggest that users find unsolicited or mistimed assistance especially frustrating in high-load moments. This aligns with the idea that, in many everyday contexts, the cost of a false positive (an unnecessary interruption) may outweigh the cost of a false negative (a missed opportunity to assist) \revise{from a user experience perspective}. 
\revise{However, false negatives remain relevant to proactive timing strategies, especially in urgent situations such as an impending meeting reminder or a fire hazard alert, where missing the moment to assist could have serious consequences. Our current design prioritizes relevance and importance over urgency, as it focuses on how working memory models can support nuanced assistive timing. In practice, urgent messages may override WM-based timing by necessity.}

\revise{To explore cognitively aligned timing without confounding urgency, we selected open-ended and non-urgent tasks that naturally engage WM. As such, we did not define a ground truth set of true positive assistance, since participants could complete tasks in multiple valid ways. This made systematic false negative tracking difficult. However, incorporating user-initiated feedback could enable future systems to capture such missed opportunities more effectively.}

\revise{The ability to report or detect false negatives would also allow future systems to better balance relevance, timing, and urgency. For instance, urgency could be inferred using heuristics-based rules that prioritize key values like safety, or LLM-based reasoning, refined through user feedback. Adaptive thresholds or user-configurable tolerances could further help modulate this balance, improving both timing effectiveness and environmental validity.}

\revise{Our findings also highlight important design tradeoffs between different approaches to modeling cognitive state. Prior systems have used physiological signals—such as heart rate variability, electrodermal activity, or respiration—to estimate cognitive load and receptivity~\cite{Sarker2014AssessingTA, Chan2020PromptoIR}. These methods can offer high-resolution data, but often require specialized hardware and raise additional concerns about privacy, interpretability, and wearability.}

\revise{In contrast, WM-based modeling provides a lightweight, explainable alternative that leverages observable behavioral cues like object interaction and conversation context. While less precise than biosensing, it offers greater transparency and easier integration with consumer-grade devices. Designers may consider combining both approaches: using physiological signals for early detection of load states, and WM modeling for timing assistance based on task semantics and interaction context.}

In summary, WM modeling presents a promising foundation for building assistants that are more aligned with how users think and feel, but must be developed with careful consideration of when not to speak, as much as when to help.

\subsection{Limitations}

While our results demonstrate the potential of working memory (WM) modeling for proactive assistance, our current implementation presents several limitations that inform future directions.

First, the system is constrained by its sensing modalities. \sys{} relies exclusively on visual and auditory signals to infer the user’s mental state. This limits its ability to capture other important data that could inform aspects of cognition, such as eye gaze, hand gesture, or tactile memory \cite{Chan2020PromptoIR}. Additionally, the current implementation does not distinguish well between first-person and third-person perspectives—for example, it occasionally encoded information triggered by the experimenter’s actions rather than the user’s own. These modality and perspective limitations underscore the need for more robust, multi-modal grounding in future systems.

Second, while our WM model is inspired by well-established psychological theories, it remains a simplified and operational approximation. Cognitive psychologists continue to debate the precise structure, encoding mechanisms, and dynamics of working memory. Rather than claiming to replicate a ground-truth cognitive model, our system offers a proof-of-concept: that WM-inspired constructs like recency, interference, and binding can serve as useful measures for timing decisions in real-world environments.

A third challenge is the coupling of assistance quality and timing in our evaluation. Even when timing is cognitively aligned, low-quality or irrelevant assistance can still feel interruptive. This coupling made it difficult to isolate the benefits of timing alone, especially in cases where the LLM-generated content lacked sufficient grounding in user goals or misunderstood context. Participants noted that mistimed or low-value messages reduced the overall sense of intelligence and usefulness, regardless of when they appeared.
That said, our core contribution is not to produce the best content for proactive assistance, but to explore how a user’s working memory state can be leveraged to strategically time such assistance. We view WM-based timing as a distinct layer that can be integrated with more goal detection and task inference mechanisms to support richer, more helpful user experiences.

\revise{Finally, our evaluation was limited to tabletop multistep tasks. These tasks were chosen to balance realism with control and to support natural WM loading while maintaining experimental consistency. They reflect real-world activities like setting a table or packing a bag, in contrast to traditional WM experiments involving memorization and recall of abstract numbers and shapes. However, we acknowledge that mobile, outdoor, or more dynamic environments may introduce additional challenges. We believe our WM-based timing model could extend to these contexts with adaptations to accommodate motion, shifting attention, and variable sensor input.}

\subsection{Future Work}

Our study opens several promising directions for future research. First, participants highlighted the importance of personalization and adaptability. Future iterations of \sys{} could learn individual user preferences over time—such as interruption tolerance, habitual forgetfulness, or common task routines—by tuning importance scores or timing thresholds dynamically. Integrating long-term memory (LTM) representations may also help contextualize working memory content with a user’s history, enabling richer inferences about task relevance and assistance value.

In addition, participants expressed a desire for more transparent and interactive feedback loops. Currently, the system operates unilaterally, with no direct channel for user correction or affirmation. Lightweight feedback mechanisms—such as confirming helpfulness, deferring suggestions, or providing “not now” options—could allow users to shape the assistant’s behavior and improve its learning over time.

\revise{Exposing the system’s internal state—such as what it currently holds in working memory or how it evaluates timing utility—could further support co-adaptation, where users develop accurate mental models of the assistant. This transparency may help users better interpret system behavior, foster trust, and modulate their interaction patterns accordingly. Additionally, future work should explore evaluating the contributions of individual components in the WM-based utility function (e.g., recency, interference, relevance) via ablation studies or parameter sensitivity analyses to better understand their impact on timing decisions.}

We also see potential in deploying WM-based timing across a broader range of devices beyond smart glasses, such as smartwatches, earbuds, or desktop companions. Each device brings different affordances and constraints for sensing and feedback, and adapting the WM model accordingly will be an important step toward more pervasive cognitive-aware systems.

While our system currently relies on audio and visual inputs, future versions should explore richer multi-modal signals. Eye gaze, hand interaction, head pose, and task-object proximity could all offer valuable cues for assessing WM load and attentional focus. Similarly, proactive assistance can expand beyond voice messages to include tactile cues (e.g., smartwatch vibrations), interface prompts, or contextual sound cues that vary in intensity or modality based on mental availability.
\revise{To further evaluate the WM-based modeling approach, future systems can implement additional adaptations to motion, shifting attention, and variable sensor input in a more dynamic setting (e.g., outdoor, participant is moving).}

Overall, our approach offers a proof-of-concept for using working memory as a foundation for timing proactive support. Future systems should expand this foundation by integrating cross-device coordination, richer multi-modal sensing, adaptive feedback, and collaborative learning to move toward more intuitive and human-aligned intelligent assistance.

\section{Conclusion}
We presented \sys{}, a proactive wearable assistant that models the user's working memory (WM) to inform the timing of just-in-time assistance. Grounded in cognitive psychology theories, our system encodes multi-modal sensor data into structured memory representations, enabling real-time reasoning about mental availability. By balancing the value of delivering assistance with the cognitive cost of interruption, \sys{} aims to provide support that aligns with the user's moment-to-moment mental state.
In our user study, \sys{} delivered fewer but more selective interventions compared to an LLM-based baseline, while yielding higher levels of user engagement. Participants reported lower \revise{frustration} levels and highlighted how assistance felt more aligned with their cognitive context. These findings suggest that WM modeling offers a promising framework for designing attentive, user-aware systems.
This work explores opportunities for integrating cognitive state modeling into wearable assistants, and highlights the importance of timing as a key dimension of proactive interaction design.

\bibliographystyle{ACM-Reference-Format}
\bibliography{references}

\appendix
\newpage
\onecolumn
\lstset{basicstyle=\ttfamily\footnotesize,breaklines=true,breakindent=0pt,breakatwhitespace=true,frame=single}
\section{Appendix}
\subsection{LLM Prompts}
\label{appendix:prompts}

\subsubsection{System Condition: WM-Aware Assistant Prompt}
\label{appendix:promemassist-system-prompt}
\leavevmode
\begin{lstlisting}
You are an intelligent assistant designed to optimize the timing of interruptions to deliver important information to users. The user is wearing a smart glasses. The smart glasses are capturing the user's ego-centric view and sounds.
Your goal is to model the user's working memory in their current task, generate potential assistance, and produce a timing decision to inform the user based on the principle of maximizing the value of interruptions while minimizing their cost.
The user will be involved in the following tasks where they interact with objects in their environment: Setting up a dining table for a meal, organizing an office desk for work, packing for a conference trip, and organizing a living room to welcome guests.
You have access to the user's Working Memory Model. The WM model is constructed by two components: perception memory and episodic buffer.
Perception memory is a list with 7 MemoryItem slots, each containing a memory item (e.g., visual image of an object, phonological words or phrase). Each MemoryItem has the following metrics: recency, relevance, and importance. Each metric is a float number between 0 and 1.
The episodic buffer consists of 4 MemoryChunks, which are groups of related MemoryItems.
You are responsible for triaging the importance scores for each MemoryItem and potential interruption messages. The recency and relevance scores are provided by the WM model.
Your objective is to determine the optimal time to interrupt the user with a new message, taking into account the recency, relevance and importance of the message, as well as the potential costs of displacement and interference to the existing working memory.
Use your knowledge of the Working Memory Model to make decisions about when to interrupt the user. Your actions should be guided by the following objectives:
Maximize the value of interruptions
Minimize the cost of interruptions
Respect the user's current task context and mental state
\end{lstlisting}

\subsubsection{Baseline Condition: LLM Assistant Prompt}
\leavevmode
\begin{lstlisting}
You are an intelligent assistant designed to deliver timely and important information to assist users in their tasks. The user is wearing smart glasses. The smart glasses are capturing the user's ego-centric view and sounds.
Your goal is to evaluate the user's current context, identify whether assistance needs to be generated, and if necessary, assist the user based on the principle of delivering timely and relevant information to aid the user in their current task.
The user will be involved in the following tasks where they interact with objects in their environment: Setting up a dining table for a meal, organizing an office desk for work, packing for a conference trip, and organizing a living room to welcome guests.
Use your knowledge of the user's context to make decisions about whether to interrupt the user.
\end{lstlisting}

\subsubsection{Task-Specific Prompts}
\leavevmode
\\
\revise{*Note that these task-specific prompts are added in both ProMemAssist and Baseline conditions during the evaluation to ensure a basic level of task understanding and assistance quality}
\begin{lstlisting}
"dining": The user is setting up a dining table for a meal with guests. They need to place objects appropriately based on categories such as eating utensils, serving dishes, and tableware.
Objects include bottle, cup, bowl, fork, spoon, orange, banana, apple, potted plant, vase

"office": The user is organizing an office space to prepare for a meeting with a colleague. They need to place objects appropriately based on categories such as electronics, reading materials, and decorations.
Objects include laptop, keyboard, mouse, cell phone, book, clock, cup, scissors, note papers, marker

"packing": The user is packing their luggage for a business trip. They need to pack objects appropriately based on categories such as clothing, electronics, and personal items.
Objects include backpack, umbrella, tie, handbag, toothbrush, laptop, bottle, banana, sunglasses, medication

"living": The user is organizing a living room to welcome guests. They need to place objects appropriately based on categories such as seating, entertainment, and decor.
Objects include remote, vase, sports ball, laptop, book, potted plant, candle, cup, snack, teapot
\end{lstlisting}

\subsubsection{Generate Episode Summary from Memory Items}
\label{appendix:prompts-episode}
\leavevmode
\begin{lstlisting}
Generate a short sentence that captures the essence of the following memory items, which are related to a specific task or activity. The sentence should be concise and descriptive, summarizing the key elements of the memory items.
Use the original task context to guide the generation of the episode. Do not assume new task context or environmental information that is not explicitly stated in the memory items.

Examples:
Memory items: [visual memory of a fork, a spoon, a bowl, phonological memory of a conversation about dinner plans]
Episode: "The user is setting up the kitchen table for dinner."
Memory items: [visual memory of a book, a laptop, phonological memory of a comment about the book's content]
Episode: "The user is discussing a book in their office space."

Output format:
{
"episode": "[short sentence capturing the essence of the memory items]"
}
\end{lstlisting}

\subsubsection{Generate Importance Scores for WM Content}
\label{appendix:prompts-importance}
\leavevmode
\begin{lstlisting}
Generate numerical importance scores for all memory items and episodes, from the range of [0,1]. Only generate the scores, do not output anything else.

Examples:
1. If the memory is task-relevant and requires immediate attention, it's very important and should have a score close to 1.
Example: The user is holding a hot cup of coffee near a child. Importance score: 0.9
2. If the memory is task-irrelevant, it should have a low importance score close to 0.
Example: The user hears a notification from their phone while working on a project. Importance score: 0.1
3. If the memory is indirectly related to the task, it should have a moderate to high importance score.
Example: The user is packing for a trip, and is reminded of the weather. Importance score: 0.65

Only output the JSON object:
{
    "perception_memory": [importance_value1, importance_value2,...],
    "episodic_buffer": [importance_value1, importance_value2,...]
}
\end{lstlisting}

\subsubsection{Generate Assistance Messages Based on WM State}
\label{appendix:prompts-sys-generate-assistance}
\leavevmode
\begin{lstlisting}
Generate a list of assistance voice messages based on the new memory item and the updated state of the working memory.
Each assistance message is one sentence providing help or reminders to the user with their current task, along with an importance score from the range [0,1]. Only generate the list, do not output anything else.

Only output the JSON object in the format below (arranged by importance):
{
  "assistance_messages": [
    {
      "message": "voice_message1",
      "importance": importance_value1
    },
    ...
  ]
}
\end{lstlisting}

\subsubsection{Baseline Assistant Assistance Generation}
\label{appendix:prompts-baseline-generate-assistance}
\leavevmode
\begin{lstlisting}
Evaluate whether to generate an assistance voice message based on the new information updated from the smart glasses: {information}
If neceesary, the generated assistance message should be a single sentence providing help or reminders to the user with their current task.

Avoid generating assistance if:
- The value and usefulness of the assistance to the user at the current action are low.
- The information is repetitive or the user is already aware of the information when performing the task.
- The assistance is not important and is more interruptive than useful to the user.

Just output the one sentence assistance if it's valuable for the user's task, and nothing else. If no assistance is needed, just return NO ASSISTANCE.
\end{lstlisting}

\subsection{Recruitment and Study Procedure}
\label{appendix:study-procedure}

\revise{Participants were recruited to complete a task simulating real-world object organization, such as setting a table or packing a bag, while wearing smart glasses.
Upon arrival, participants signed a consent form and received a brief overview of the study. The introduction statement is as follows: ``You are wearing smart glasses and performing tasks that involve interacting with physical objects in different scenarios. You will also engage in conversation with other people (the experimenter) in the scenario. You will receive AI assistance throughout the session. The focus of this study is on the timeliness of the AI assistance, whether the assistance timing felt just-in-time or interruptive.''}

\revise{For each of four task scenarios, the participant was given contextual instructions before starting (e.g., you are setting up your new office desk next to your co-worker). During the task, the experimenter introduced new information—some relevant to the participant’s activity, others intentionally irrelevant—to simulate interruptions. The AI system could provide assistance, and the experimenter marked any observed positive reactions for later discussion in the interview.}

\revise{After each task, participants completed a short questionnaire. At the end of the session, they completed an exit survey and participated in a semi-structured interview, where they reflected on their experiences and the system’s performance.}

\revise{All sessions were recorded using camera and screen capture tools, and all system conditions were kept blind to participants until after the interview and debriefing.}

\subsection{Task Setup and Materials}
\label{appendix:task-details}

Each task round was designed with a consistent structure: the participant interacted with a set of physical objects (10 distinct object types for each task) while receiving intermittent spoken prompts from the experimenter. These prompts included both task-relevant and irrelevant information to simulate naturalistic interruptions. 
% Tasks lasted ~5 minutes each, and participants filled out post-task surveys afterward.

\subsubsection*{Task 1: Dining Setup}
\textbf{Objects:} bottle, cup, bowl, fork, spoon, orange, banana, apple, potted plant, vase

\textbf{Task Goal:} Set up a dining table to welcome guests for dinner.

\textbf{Scripted Prompts and Potential Follow-up Action:}
\begin{itemize}
    \item “I need to make a fruit salad with apples and bananas for dessert.” $\rightarrow$ Put apples/bananas aside
    \item “The spoon is for the coffee in case anyone needs stirring.” $\rightarrow$ Put spoon near cup
    \item “The potted plant needs more sunlight.” $\rightarrow$ Move plant toward window
    \item \textit{“I'm thinking of redecorating the living room...”} [irrelevant]
\end{itemize}

\subsubsection*{Task 2: Office Organization}
\textbf{Objects:} laptop, keyboard, mouse, cell phone, book, clock, cup, scissors, notepaper, marker

\textbf{Task Goal:} Organize a new office desk for a meeting.

\textbf{Scripted Prompts:}
\begin{itemize}
    \item “I have a meeting soon and need to review notes.” $\rightarrow$ Keep notes accessible
    \item “The clock is not working properly.” $\rightarrow$ Put clock away
    \item “I use that blue book every day for reference.” $\rightarrow$ Place book nearby
    \item \textit{“We have a charity event next month.”} [irrelevant]
\end{itemize}

\subsubsection*{Task 3: Packing for a Trip}
\textbf{Objects:} backpack, umbrella, tie, handbag, toothbrush, laptop, bottle, banana, sunglasses, medication

\textbf{Task Goal:} Pack a backpack for a business trip

\textbf{Scripted Prompts:}
\begin{itemize}
    \item “The weather forecast says it’s not gonna rain.” $\rightarrow$ Skip umbrella
    \item “Make sure you have snacks for the road.” $\rightarrow$ Pack fruits/snacks
    \item “The conference is business formal.” $\rightarrow$ Pack tie
    \item \textit{“I want to try a new coffee shop in the city.”} [irrelevant]
\end{itemize}

\subsubsection*{Task 4: Organizing Living Room}
\textbf{Objects:} remote, vase, sports ball, laptop, book, potted plant, candle, cup, snack, teapot

\textbf{Task Goal:} Style a living room space to host guests.

\textbf{Scripted Prompts:}
\begin{itemize}
    \item “The guest’s child loves basketball.” $\rightarrow$ Leave ball accessible
    \item “It’s daylight so we don’t need candles yet.” $\rightarrow$ Remove candles
    \item “Let’s do movie night later—I can set up the TV.” $\rightarrow$ Retrieve remote
    \item \textit{“The neighbors are having a party tonight.”} [irrelevant]
\end{itemize}

\subsection{Interview Questions}
\label{appendix:interview questions}
\begin{itemize}
    \item \revise{Can you tell me about a time when the system provided assistance at a moment when you really needed it? How did that make you feel?}
    \item \revise{Were there any times when the system provided assistance at an undesirable timing? How did that affect your experience?}
    \item \revise{Were there any times when the system provided information that was useful to you? How did you feel about receiving that information?}
    \item \revise{Can you think of a time when the system provided incorrect or irrelevant information? How did you handle that situation?}
    \item \revise{Can you describe a time when you felt fully engaged and focused during your task while using the system? What were you doing during that time?}
    \item \revise{Were there any times when you felt distracted or disengaged from the task? What do you think caused that feeling?}
    \item \revise{How do you feel about your sense of control and agency?}
    \item \revise{How much do you trust the system?}
\end{itemize}

\end{document}